# Comment on "Feynman Effective Classical Potential in the Schrodinger Formulation"

In a recent Letter [1] the zero temperature limit of the effective classical potential (ECP) is obtained through the minimization of $<\psi|\hat{H}|\psi>$ under the constraint $<\psi|\hat{x}|\psi> = x$, and those states that realize this minimization are called minimal energy wave packets (MEWP). The authors approximate the time evolution of a wave packet, initially in an MEWP configuration, with the classical motion of a particle in the ECP. To build up their approximation, however, *they incorrectly resort to the Ehrenfest classical limit*. Below, we point out *the proper variational principle for the effective action* (EA) which indeed underlies the actually *highly non-classical approximation* considered in [1] (where the authors describe the tunnelling of a particle in a double well potential), and we show the difference with a truly Ehrenfest-like motion. Once the connection with the EA is established, we can not only correctly understand the approximation in [1] as the lowest order approximation of a systematic expansion, but we can also improve on it. In addition we note that the zero temperature ECP is nothing but the well known quantum mechanical effective potential (EP) and that the minimization procedure used in [1] was actually established long ago and can be found in many textbooks [2].

The example studied in [1] is the tunnelling in a double well $V_{dp}(x) = -x^2/2 + \lambda x^4/24$. The dynamical evolution of a wave packet (WP) in $V_{dp}$ can be analyzed [3] via the EA, $S_{eff}[x(t)]$, whose static limit is the EP, $V_{eff}(x)$, i.e., the ECP of [1]. A variational principle that allows one to determine the EA as the stationary, time integrated matrix element of $i\partial_t - \hat{H}$ between time dependent states subject to a double constraint has been derived in [4]. Remarkably, when restricting onself to the diagonal matrix element of $i\partial_t - \hat{H}$ in a state $\psi$, the argument $x(t)$ of the functional $S_{eff}[x(t)]$ has the meaning of the time dependent coordinate expectation value in the state $\psi$ [4]. Let us consider the derivative expansion of the EA, $S_{eff}[x(t)] = \int dt(-V_{eff}(x) + Z_{eff}(x)(\partial_t x)^2/2 + Y_{eff}(x)(\partial_t x)^4/24 + \cdots)$. One immediately realizes that in the lowest order approximation, i.e. $Z_{eff}(x) = 1$, $Y_{eff}(x) = \cdots = 0$, the dynamical evolution of the WP is approximated by $V_{eff}$, which is what has been done in [1]. This provides the *only correct framework* to interpret the results in [1]. This is nothing but the lowest order of a systematic approximation and we know how to improve on this result. We make use of the Wilsonian action renormalization group flow equation [5] to determine $V_{eff}$ and $Z_{eff}$ by solving numerically two coupled partial differential equations [6]. In Fig.1, a $x$-$v(=\dot{x})$ diagram for the WP time evolution (dotted line) with $\lambda = 6$ is compared to the ones relative to the functions $x(t)$ which extremize $S_{eff}$ either with $Z_{eff} = 1$ as in [1] (dashed line), or including $Z_{eff}(x)$ (solid line). Our WP at $t = 0$ is a gaussian centered at x=0.7 (for comparison with [1]) whose width is fixed by the curvature $V''_{eff}(x = 0.7)$. Note in Fig.1 the improvement of some percent that is obtained when $Z_{eff}(x)$ is included.

To conclude we stress here that the Ehrenfest theorem establishes, under "classical" conditions *which do not include the ones considered in [1]*, the equation of motion of $x(t)$ in terms of $V_{dp}$, not of $V_{eff}$. As a good example of truly Ehrenfest classical motion, we present here the motion of a WP in a deep double well potential ($\lambda = 0.1$), *around one of the classical minima*. The motion of the mean position of the WP ( solid line in Fig.2) is reasonably well approximated by the Ehrenfest motion of $x(t)$ (dashed line).

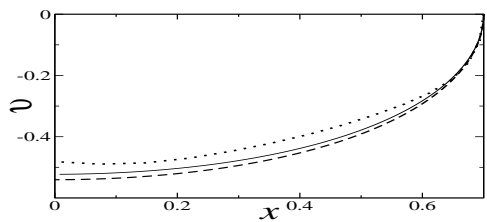

FIG. 1. Phase space trajectories with $\lambda = 6$ in $V_{dp}$.

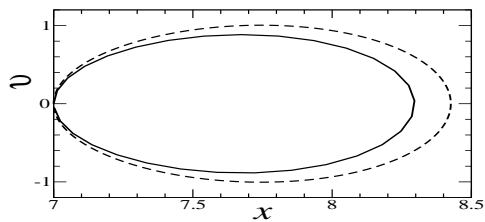

FIG. 2. Phase space trajectories with $\lambda = 0.1$ in $V_{dp}$.

G. Andronico,[1] V. Branchina,[2] and D. Zappalà[1]
[1]INFN, Sezione di Catania,
Corso Italia 57, Catania, 95125, Italy
[2]IReS, 23 rue du Loess BP28, 67037 Strasbourg cedex 2, France

1